\documentclass[twocolumn,pacs,preprintnumbers,amsmath,amssymb]{revtex4}
\usepackage{graphicx}
\usepackage{dcolumn}
\usepackage{bm}
\begin{document}
\voffset 0.5 truein

\preprint{}

\title{ ECoG observations of  power-law scaling in the human  cortex }
\author{Kai J. Miller$^{1}$, Larry B. Sorensen$^{1}$, Jeffrey G.~Ojemann$^{2}$, and Marcel den Nijs$^{1}$ }

\affiliation{Departments of Physics$^{1}$   and  and Neurological Surgery$^{2}$ , 
University of Washington, Seattle, Washington 98195, U.S.A.}
\date{\today}

\begin{abstract}

We report the results of our search for power-law electrical signals in
the  human brain, using subdural electrocorticographic recordings from the surface of the cortex.  
The power spectral density (PSD) of these signals has the power-law form $ P(f)\sim  f^{-\chi} $  from 80 to 500~Hz.  
This scaling index  $\chi = 4.0\pm 0.1$ is universal, across  subjects, area in the cortex, and local neural activity levels.
The shape of the PSD does  not change with local cortex activity, only its amplitude increases.
We observe a knee in the spectra at $f_0\simeq 70$~Hz, implying  the existence of a characteristic time scale
$\tau=(2\pi f_0)^{-1}\simeq 2-4$~msec.
For $f<f_0$ we find evidence for a power-law with $\chi_L\simeq 2.0\pm 0.4$. 
\end{abstract}

\maketitle

The human brain is arguably the most complex structure known to mankind
and on the verge of starting to grasp its own  inner workings.
How do our brains compute?  How fast do they  compute?  How do they store information?
How universal is all of the above?
Ever since the first electroencephalography (EEG) recordings  in 1924, the study of the 
electrical activity of the human brain has focused on its prominent low-frequency features,
in particular the excitatory and inhibitory rhythms at specific frequencies, like the 
 $\alpha$ (10~Hz) and  $\beta$ (20~Hz)  rhythms \cite{eeg}.
Traditional EEG studies are limited to $f<100$~Hz. 
The fundamental processes at the individual neuron scale 
suggest a role of higher frequencies:
the time of flight of a spike along an axon, the synaptic neuro-transmittor diffusion time, the integration time of the dendritic arbor. 
They are all near or sub 10~ms \cite{timescales}.
Synchronization and correlations associated with them are expected to exist at 
least up to 1~kHz.

\begin{figure}[b]
\centering
\includegraphics[width=78 mm]{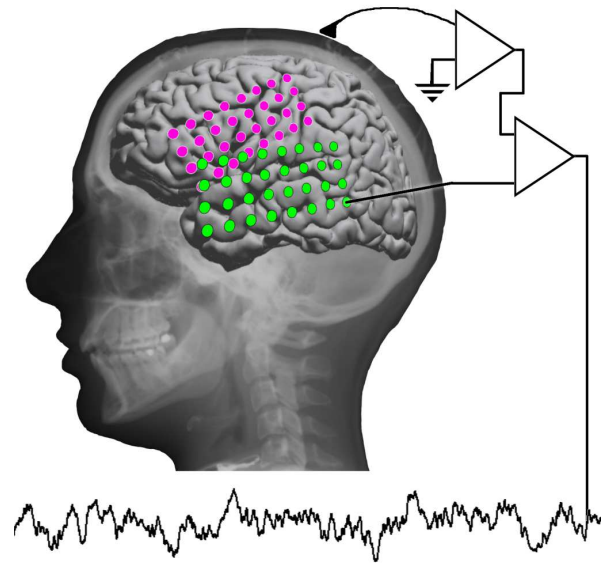}
\caption{The electrode array locations are shown on a template brain for subject 1 (S1 - purple, temporal) and subject 2 (S2 - green, fronto-parietal).  Potentials of all 32 channels are measured simultaneously 
with respect to a scalp  reference and ground before pairwise re-referencing.}
\label{fig:schematic}
\end{figure}

Electrocorticographic (ECoG) recordings from the subdural surface of the cortex have recently made it possible to examine the electrical activity of the human neocortex with finer spatial and temporal fidelity than EEG. 
An array of electrodes is placed directly on the surface of the cortex, see Fig.~\ref{fig:schematic}.
The absence of the skull and surrounding tissue  increases the electrode voltage  while the close proximity to the cortex means 
that ECoG records very  local phenomena.
For example, changes in the classical $\alpha \& \beta$ rhythms appear  spatially uniform for a given set of tasks in EEG, but vary strongly spatially within the ECoG array for the same tasks \cite{eegecogspatial}.

The cortical surface potentials from sub-dural arrays reported in this study were obtained from 
20  participants receiving clinical monitoring for the localization of seizure foci prior to resection.  
Each was informed about, and consented to participate in, the University of Washington internal-review-board-approved experimental protocol. 
The voltage was sampled at 10~kHz (2 subjects) or 1~kHz  (18 subjects) using Synamps2 amplifiers 
(Compumedics-Neuroscan, San Antonio, TX) in parallel with long term monitoring
(Xltek, Oakville, ON) from 32 platinum electrodes, encased in silastic, 
in an 8x4 configuration (4~mm in diameter, 
with 2.3~mm exposed, separated by 1~cm, center-to-center, Ad-Tech, Racine, WI).  

Our earlier studies \cite{pwrlawhyp} revealed  the absence of distinct peaks in the power spectrum
beyond $f\simeq 60$~Hz.  We hypothesized the existence of a power-law of the form $P(f)\simeq A f^{-\chi}$
at these higher   frequencies, and named it the $\chi$-band/index, but
the 1~kHz sampling rate truncated the signal at these high frequencies.
The purpose of the study reported here was  to determine,  as accurately as possible, whether there is 
indeed such a power-law in the human cortical power spectrum, and how it might change with cortical activity (universality),
by using a higher, 10~kHz sampling rate.

Power-laws represent scale free behavior, the finding of which immediately evokes scale free networks, 
complexity, avalanches, and self-organized criticality (SOC).  Unfortunately, many such networks and 
processes  are not large enough or can not  be monitored precisely long enough  to establish the
scale invariance  convincingly  \cite{otherpwrexs, critphenomena}.
The human brain is arguably the most  complex and largest  network
available and may provide the best
opportunity  to observe scale free behavior in a natural setting.
Each ECoG electrode measures the voltage power spectral density
from a specific cortical surface area associated with a specific set of  functions.
An  electrode pair probes about $10^6$ neurons, and each neuron has up to $10^4$ synapses \cite{timescales}. 
This has  not gone unnoticed, and the neuroscience literature is
awash with attempts to  interpret experimental low-frequency data with scale free concepts and models;
with only limited  success and leaving many questions \cite{neuropwrexs}.
In this paper we  firmly establish the existence of the power-law in the $\chi$-band 
and the actual  value of the scaling index,  $\chi$.
We obtain remarkable  accuracy, particularly  compared to many  
recent  studies of power-law phenomena in nature \cite{otherpwrexs}.   
Our accuracy approaches  that of what was required and customary  in equilibrium critical  phenomena \cite{critphenomena}. 

 \begin{figure}[t]
\centering
\includegraphics[width=92 mm]{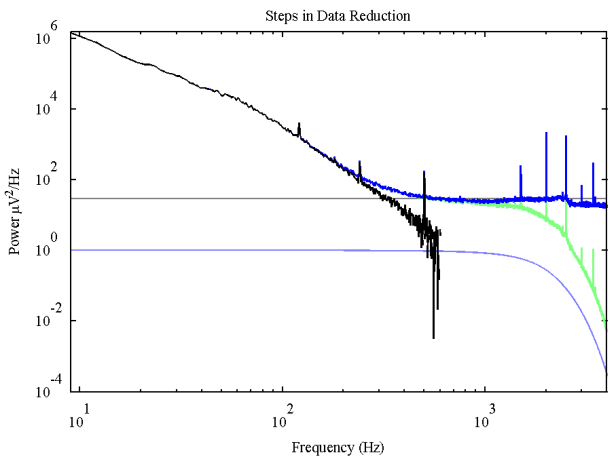}
\caption{Illustration of the steps taken to remove  the amplifier 
roll-off and noise floors from the raw power spectra (green to dark blue to black)
The amplifier built-in amplitude attenuation was determined by scanning through 15-4000~Hz sine waves 
and fitting a smooth function to the attenuation at each frequency (light blue).  
A characteristic distribution of amplifier noise floors  (current and potential noise) was generated by 
measuring the potential across an equivalent conformation of resistors (grey).
The sharp line noise spikes at 60~Hz and its harmonics were  excluded in our analysis.}
\label{fig:spcorr}
\end{figure}

We characterized the power-law in the cortical spectrum during a simple fixation task.  
The subjects fixated on an ``$\times$" on the hospital room wall 3~m away from the bed for 
130~s (subject S1) or 190~s (subject S2), with their eyes open.  
The time-dependent voltage between each electrode and the reference electrode was 
measured at $10~kHz$, digitized, and stored for spectral analysis.  To further reduce the common mode 
noise from the environment, the digitized electrode voltages were converted digitally into a
set of voltage differences between each near-neighbor pair of electrodes.  This significantly 
removed the high-frequency common mode noise and make it possible to measure well 
above 100~Hz.  For our 4 by 8 array, the 32 individual electrode voltages produce 52 near- 
neighbor voltages (Fig.~\ref{fig:schematic}).  All of the measurements reported in this Letter are for  near-neighbor voltages. 

We carefully characterized the amplifiers, their low pass filtering  and their noise floor.
These external factors affect  the power spectrum measurement dramatically.
They mask and obscure the underlying power-law from the brain signal.
Demonstration of the power-law would not have been feasible without these corrections. 
Fig.~\ref{fig:spcorr} shows these corrective steps in the data reduction.  
First, the power spectral density is calculated from the Fourier transform of the time-varying near-neighbor voltages in 1~s
Hann-windowed epochs, overlapping  by 0.5~s.  
These are averaged into  uncalibrated spectra
(marked green in Fig.~\ref{fig:spcorr}). They  suggest a power-law shape, but the roll-off, apparent above  1~kHz, masks it.
This roll-off  does  not represent a characteristic  high-frequency in the brain. It originates in the amplifiers.
We measured the gain versus frequency of the amplifiers independently.
Their spectral bandwidth  shape follows a classic low-pass filter (shown as  blue in Fig.~\ref{fig:spcorr}).
Removing the  amplifier response from each uncalibrated spectrum leads to  
spectra  (dark blue) that  level-off at high frequencies, indicating that we hit a noise floor.
This noise floor does not reside in the cortex  either.
We measured  the amplifier input noise for our amplifier-digitizer system independently.
It is  of order  $\simeq 4~\mu$V per root Hz for every amplifier.
Subtracting this from  the spectrum 
leads to a  power spectrum that remarkably tightly fits a straight line in the log-log plot 
(the black line see Fig.~\ref{fig:spcorr})
until at  500~Hz  the signal disappears into the fluctuations of  the amplifier noise floor.
Future experiments with lower noise floor amplifiers will tell how high in frequency the power-law actually continues.
 
\begin{figure}[b]
\centering
\includegraphics[width=86mm]{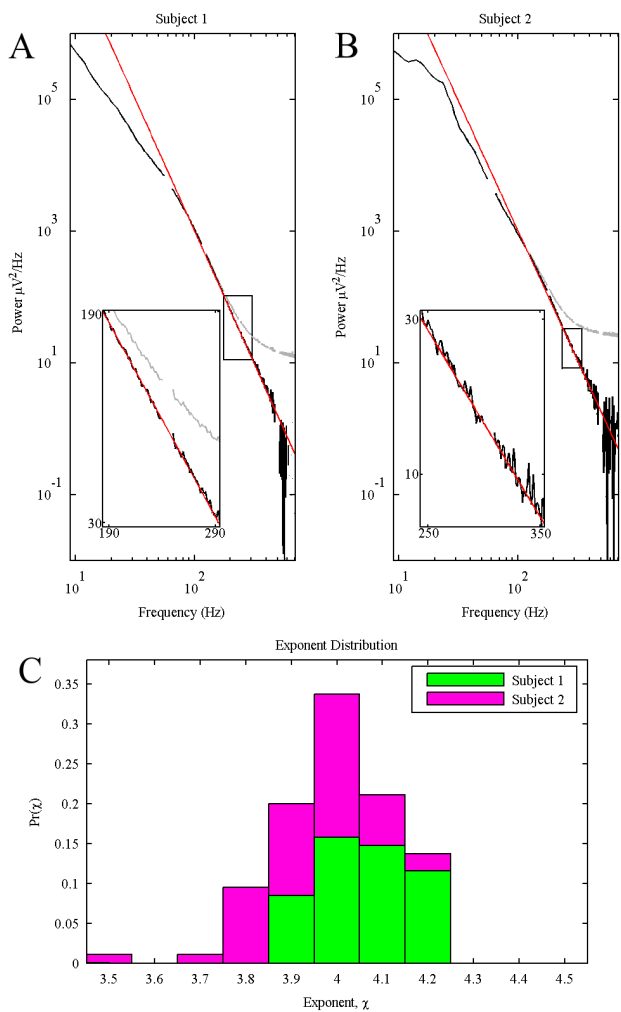}
\caption{(A)-(B): Averaged electrode-pair power spectra (black) for subjects S1 and S2.
The red lines, the best power-law fits (see text) with $\chi=4.03$ (S1) and  $\chi$=3.96 (S2),  
fit the data very well from 80-500 Hz.  The spectrum before noise floor subtraction is shown in gray. The S1 spectrum has a  knee at 
at  $f_0=70$~Hz and appears linear below $f_0$.
(C): histogram of power-law fits  from  individual electrode-pair spectra, with mean values $\chi=4.06$ (S1; green) and $\chi=3.94$ (S2 purple).}
\label{fig:fitfig}
\end{figure}  

Small variations and uncertainties  
in the amplifier noise floor $C$ significantly affect the quantitative analysis  of the  power-law.
For that reason we perform 3 parameter fits, $P(f)= A f^{-\chi} + C$.
All such estimates for $C$  are  within the uncertainty  of  our independent amplifier noise floor measurements.

Fig.~\ref{fig:fitfig}  shows the combined spectrum, averaged over electrode pairs, for  each subject.
We exclude electrode pairs where one of the  electrodes sits primarily on top of a vasculature, 
resulting in a much lower power in the signal and increased sensitivity to noise
(4 channels for S1, 5 channels for S2).
The inserts  illustrate the quality of the power-law, the  jitter around the average curve is  more than  one decade  down
from the signal, for all $f<200$~Hz, beyond which the amplifier  noise floor  fluctuations start to kick-in.
The exponent  $\chi$  and the parameters $A$ and $C$ 
are estimated via a set of  log-log type least-squares linear fits of the power spectral density
between 80 and 400~Hz, excluding harmonics of 60~Hz. 
This leads to  $\chi= 4.03\pm 0.1$ for S1 and $\chi=3.96\pm 0.1$ for S2.
The error bars are based on robustness against range shrinking as well as 
the deviations of the best fit  with respect to the actual data
across  the entire frequency range $80<f<400$~Hz.

To test for universality,  we also performed the exact same type of fits to 
each individual electrode pair spectrum.
The histograms  for S1 and S2 in Fig.~\ref{fig:fitfig} overlap well and the 
variations within each can easily be attributed to stochastic external issues, 
such as variations in electrode-cortex distance and vasculature.
A such, we reached the accuracy limit  imposed by  the experimental set-up.
Within this accuracy, $\chi$ is universal; it does not vary  with subject nor  specific brain areas.  
The mean value of the histogram exponents for S1 is $\chi=4.06$ 
(STD=0.10, N=48), and for S2 $\chi=3.94$ (STD=0.13, N=47),
consistent with the above fitting analysis on  the two averaged  spectra.
We conclude that  $\chi = 4.0\pm 0.1$ throughout  the frequency range $80<f<400$~Hz.
   
Previous estimates of a power-law in the cortical spectrum \cite{cortfit}  focused on low frequencies.
The averaged power spectrum of S1, in Fig.~\ref{fig:fitfig} shows a knee at $f _0\simeq 70$~Hz, 
and suggests a different power-law below $f_0$. Both seem absent in S2.
The $\alpha \& \beta$ rhythms are strongly pronounced in every channel pair of S2 
(and are clearly visible in Fig.~\ref{fig:fitfig}). 
They obscure whatever power-law might be present  underneath.
Eight electrode pairs from S1 lack  $\alpha \& \beta$ rhythms. 
They  may be absent in these local cortical areas, or
so  tightly synchronized that they cancel out  in the electrode pair voltage  difference.
A  simple minded  $P(f)\simeq A f^{-\chi_{L}} $ fit through  
their average from 15-80~Hz, yields  $\chi_L= 2.57\pm 0.15$ (N=8). 
However, such local fits are inherently dangerous.
They are blind to the  global properties of the spectrum.
Our  high frequency analysis  of $\chi $  is already a  clear illustration of this.
The amplifier roll-off and noise floors truncate our data only beyond $f\simeq 400$ ~Hz,
but  they affect the spectrum already at much lower frequencies (Fig.~\ref{fig:spcorr}).
Similarly,  the high frequency $\chi=4.0$ power-law is already in-play below $f_0$. 
Indeed, the global two-powerlaws form  

\medskip
\centerline{
$
P(f) \sim A \frac{1}{ 1+ \left( \frac{f}{f_L } \right)^{\chi_{L}} }
                   \frac{1}{1 + \left( \frac{f}{f_0 } \right)^{\chi_H} }
$
} \hfill\break
with constraint $\chi_L+\chi_H=\chi=4$,  shifts $\chi_L$ down to $\chi_L=1.99$ (STD=0.21, N=8), with cutoff $f_L\ll 15$~Hz.

The 8 S1 channels  form  a dangerously  small ensemble.
We performed the same analysis on  1~kHz data for the exact same 2 minute fixation task from 16 subjects,
with electrode arrays in the lateral frontal/temporal/parietal cortex.
The spectra were determined and corrected for amplifier-digitizer frequency dependent attenuation in the same way as the 
10~kHz data. We selected again only  those channel pairs (N=116) that lacked  the $\alpha \& \beta$ rhythms.
Naive fits  from 15-80~Hz (excluding 57-63~Hz line noise) of these corrected spectra 
yield again  $\chi_{L}=2.5$ (STD=0.4, N=116);  while  the same 
global two power-law form, shifts this down again to $\chi_{L}=2.0$ (STD=0.4, N=116)
when   setting $f_L=1$ and $f_0 =70$. 

\begin{figure}[t]
\centering
\includegraphics[width=82 mm]{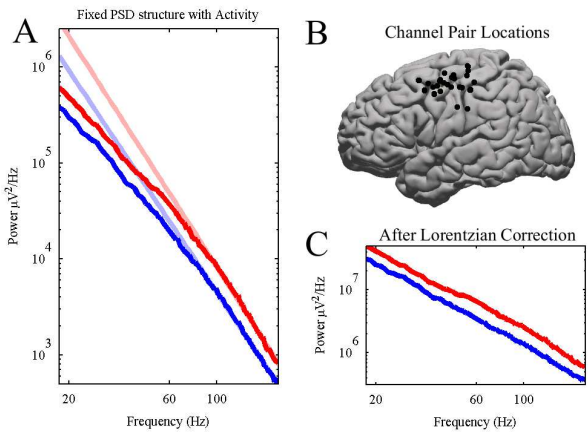}
\caption{The average shift in power spectral density in electrodes  after the PCA decomposition 
in the hand cortical area during finger movement, for subjects 3-8; 5 electrode-pair channels each (30 total).  
(A)  demonstrates that  movement (red) increases the  overall power by a factor  1.8 while preserving the shape of the rest (blue) spectrum.
$f>f_0$ power-law fits (grey) are consistent with $\chi=4$.
(B) Electrode pair channel locations (interpolated)   across all subjects, projected to the left-hand side. 
(C) Remaining spectra after dividing out a Lorentzian, $1/\left( 1 + (f/70)^2\right)$ 
to illustrate consistency with the two-powerlaws  form  with $\chi_L\simeq 2$.
}
\label{fig:shiftfig}
\end{figure}

The uncertainty  in the value of $x_L$  remains quite large. Even the  existence of this low frequency power-law remains 
in question, until we  collect more 10~kHz data. The  analysis  also requires a reliable decomposition
of the low frequency  $\alpha \& \beta$ rhythms from the background. We demonstrated  that principle component analysis 
(CPA) techniques can achieve this \cite{Kai-next, pwrlawhyp}.  
Fig.~\ref{fig:shiftfig} illustrates this for 1~kHz data.
Five  subjects performed a finger movement task.  
A visual cue indicated when to move a given finger (repeatedly flex/extend one finger at a time; 
opposite side of body from grid placement), and the position of each finger was recorded using a dataglove (5dt; Irvine, CA).  
Samples of movement spectra of each type were calculated from the FFT of 1~s  Hann-windowed 
epochs centered at the maximum displacement of the finger during each flexion.  
Rest spectra were calculated from epochs in which there was no hand movement.  
The spectra were corrected for amplifier-digitizer frequency dependent attenuation.  
Differences  between  variations in $\alpha \&\beta$ versus high frequency 
allows a PCA 
removal of the $\alpha \&\beta$ peaks  from all the power spectra. 
We were able to identify electrode pair channels which demonstrated significant shifts in power during specific activity (single finger movement  type) versus rest.  
A subset of five such channels was chosen naively (based upon significance rather than being hand picked) from each subject, and the average spectrum for the specific 
movement type causing change (the ``active" state) was compared to  the average rest spectrum (the ``inactive" state), after the $\alpha \& \beta$ peaks had been 
decoupled and largely removed.

Fig.~\ref{fig:shiftfig} addresses several issues:
Earlier \cite{ pwrlawhyp, eegecogspatial}, we observed changes in the $\chi$-band
during behavior tasks and hypothesized  how those can be used directly to quantify activity in the brain in a variety of practical settings;
e.g.,  the total power in the $ \chi$- band increases with activity  versus rest.
Fig.~\ref{fig:shiftfig} demonstrates that this  increase in spectral power with activity extends over  all  frequencies.
Moreover it  strongly suggests that  the shape of the spectrum is preserved  (universality).
The above  $\chi_L-\chi_H$ two-powerlaws form  fits the shape well;  with 
$f_0\simeq 70$~Hz, $\chi\simeq 4$,  and $\chi_L\simeq 2$, within 10\%;
a level of accuracy to be expected within  the limitations of the 1~kHz nature of this data. 
The  active/inactive power ratio $R$ (between the amplitudes $A$)  is unlikely a universal number.
In this data set its geometric mean is  equal to $R=1.81$  with a variation of order 0.34 (N=25).

What does all of this teach us about computations and correlations in the brain?
EEG and ECoG  voltages represent  the superposition of the electric current dipole fields generated by the  very large collective
of nearby neurons and their associated  ionic channel  currents, propagated  through the complex mass of ionic extracellular liquid and neuronal and glial membranes.
This is a quite complex  phenomenon and not yet well understood quantitatively.
Moreover, the underling  neuron computational issues  remain in flux, e.g,  understanding the relative  roles of computations at 
the level of dendritic trees versus  those at larger length and time scales 
associated with the connectivity  the network of neuron.

If the value of the high-frequency  exponent $\chi=4.0\pm 0.1 $ were  distinct from an integer,
we could have reported safely to  have observed SOC type complex scale free behavior in the cortex. 
The uncertainty in  $\chi_L=2.0\pm 0.4$ still leaves room for this. Perhaps 
SOC behavior  (if it exists) is  only expressed in more subtle ways in ECoG.
In the case  $\chi_L=2$,  the spectrum is well described as a
product of two $1/(1+(f/f_c)^2 )$  Lorentzian shapes. 
These  can originate without any  SOC complexity:  such as white noise with two filters; as the
product of an exponential decaying  correlation function and a temporal form factor;  as
two processes with definite correlation times; etc.

Our results place definite constraints on future brain modeling.
ECoG spectra  definitely  scale as $P\sim f^{-\chi}$ across all  $70<f<400$~Hz.
The value $\chi =4.0\pm 0.1$ is universal, across  subjects, areas in the cortex, and local neural activity levels.
The knee in the spectra at $f_0\simeq 70$~Hz, implies the existence of a characteristic time scale
$\tau=(2\pi f_0)^{-1}\simeq 2-4$~msec, probably originating at the neuron length scale
and surviving the coarse graining to the 2~mm electrode size length scale.  

\emph{Acknowledgment --}
This research is supported by NSF grants BCS-0642848 (KM,JO)  and  DMR-0341341(MdN), 
We like to thank the patients and staff of Harborview Hospital, Seattle, WA.


\begin{references}
\bibitem{eeg} 
See e.g.,  E.~ Niedermeyer and F.L. ~da Silva, {\it Electroencephalography: Basic Principles, Clinical Applications, and Related Fields} (Lippincott Williams \& Wilkins, 2005).

\bibitem{timescales} for a review see:
G.M.~Shepherd , {\it The Synaptic Organization of the Brain}  (Oxford Univ. Press N,  1998).

\bibitem{eegecogspatial} 
Crone {\it et al.},  Brain 121:2271Ð2299 (1998) and Brain 121: 2301Ð2315 (1998).

\bibitem{pwrlawhyp} 
Miller, KJ, {\it et al.} J.~Neurosci, 27(9):2424-2432  (2007);
NeuroImage, 37:504Ð507 (2007);  IEEE-TBME, in print.

\bibitem{otherpwrexs}  
A Clauset, {\it et.al. }, {\it Power-law distributions in empirical data}, eprint arXiv: 0706.1062 (2007)

\bibitem{critphenomena} 
For a discussion see, e.g., 
Marcel den Nijs, {\it Growth in Scale Invariance}, APCTP Bulletin, {\bf 3}, 18-23 (2000). 

\bibitem{neuropwrexs} 
John M.~Beggs and Dietmar Plenz, J. Neuroscience 23(35), 11167(2003);
M.~Shimono {et. al.} PRE {\bf 75}, 051902 (2007);
M.~Usher, {\it et. al.},  PRL {\bf 74} 326 (1995). 

\bibitem{cortfit}
C. ~B\'edard, {\it et.al.}  PRL {\bf 9}, 118102 (2006);
L.~Arcangelis  {\it  et. al.}  PRL {\bf 96} ,028107 (2006).

\bibitem{Kai-next} Miller, KJ, {\it et al.}, to be published.


\end{references}
\end{document}